\pdfoutput=1
\documentclass[american]{article}
\usepackage{jinstpub}
\usepackage{amsmath,amsfonts}
\usepackage{courier}
\usepackage[utf8]{inputenc}
\usepackage{listings}
\usepackage{microtype}
\usepackage[obeyFinal,textsize=footnotesize]{todonotes}
\usepackage[todos]{CMImacros}

\newcommand{\tpxdevice}{\texttt{TimepixDevice}\xspace}
\newcommand{\pymepix}{\texttt{PymePix}\xspace} % to be discussed/defined
\newcommand{\pymepixacq}{\texttt{pymepix-acq}\xspace} % to be discussed/defined
\newcommand{\pymepixview}{\texttt{pymepixviewer}\xspace} % to be discussed/defined
\newcommand{\pipeobj}{\texttt{BasePipelineObject}\xspace}

\newcommand{\acqpipeline}{\texttt{AcquisitionPipeline}\xspace}

\begin{document}
\title{\pymepix: A python library for SPIDR readout of Timepix3}%
\author[a]{Ahmed~F.~Al-Refaie,}%
\author[a,b]{Melby~Johny,}%
\author[c]{Jonathan~Correa,}%
\author[c]{David~Pennicard,}%
\author[d,e]{Peter~Svihra,}%
\author[f]{Andrei~Nomerotski,}%
\author[a,g]{Sebastian~Trippel,}%
\author[a,b,g]{and Jochen~Küpper\,}%
\affiliation[a]{Center for Free-Electron Laser Science, Deutsches Elektronen-Synchrotron DESY,
   Notkestrasse 85, 22607 Hamburg, Germany}%
\affiliation[b]{Department of Physics, Universität Hamburg, Luruper Chaussee 149, 22761 Hamburg,
   Germany}%
\affiliation[c]{Deutsches Elektronen-Synchrotron DESY, Notkestrasse 85, 22607 Hamburg, Germany}%
\affiliation[d]{Department of Physics, Faculty of Nuclear Sciences and Physical Engineering, Czech
   Technical University, Prague 115 19, Czech Republic}%
\affiliation[e]{School of Physics and Astronomy, University of Manchester, Manchester M139PL, United
   Kingdom}%
\affiliation[f]{Physics Department, Brookhaven National Laboratory, Upton, NY 11973, USA}%
\affiliation[g]{Center for Ultrafast Imaging, Universität Hamburg, Luruper Chaussee 149, 22761
   Hamburg, Germany}%
\emailAdd{jochen.kuepper@cfel.de}
% \arxivnumber{}
% \keywords{Only keywords from JINST's keywords list please}
\date{\today}%
\abstract{\pymepix is a new Python 3 library that provides control and acquisition for the
   Timepix3-SPIDR hardware. The rich set of data-structures and intuitive routines reduces time and
   coding effort to quickly configure, acquire, and visualize data from Timepix3. The highly
   extensible high-performance data-pipeline allows for alteration of the Timepix3 datastream into a
   form that is convinient for the user. This library is intended to be easily inserted into a
   standard scientific software stack as well as to allow for more direct interaction of Timepix3
   with interactive flavors of Python. Included with the library are two example programs using
   \pymepix: \pymepixacq is a command line control and acquisition program that can capture UDP
   packets and decode them into pixels and triggers. The second is \pymepixview, an online control
   and data-acquisition program for general use, but with features geared toward mass-spectroscopy
   and ion imaging.}%
\maketitle
\flushbottom

\section{Introduction}
Timepix3 is an application specific integrated circuit (ASIC) hybrid pixel detector developed by the
MediPix3 collaboration~\cite{Poikela:JINST9:C05013} as the successor of previous versions of
Timepix~\cite{Llopart:NIMA581:485}. Each pixel in the $256\times256$ matrix acts independently and
is capable of recording hits with timing and energy information at MHz rates. Timepix3 can operate
in either frame-based mode like standard CMOS and CCDs or a sparse data-driven mode where each pixel
hit is immediately sent out as a data packet -- containing information on the pixel coordinates,
time of the hit, and energy deposited, with a deadtime of about 1~\us~\cite{Frojdh:JINST10:C01039}.
Each pixel has a customizable threshold level that determines when a hit is registered. If a signal
causes a crossing of this threshold then the hit is registered along with the time-of-arrival (ToA)
information and time taken for the signal to fall below the threshold, also referred to as the
time-over-threshold (ToT) duration. ToA of a pixel is encoded as a 14-bit value operating at 40~MHz
giving a temporal resolution of 25~ns and maximum time of 83~\us. The resolution can be refined
further using the 4-bit fast time of arrival (FToA) operating at 640~MHz improving the resolution to
1.56~ns. The ToT estimates the energy deposited into the pixel recorded as a 10-bit value at 40~MHz
giving a resolution of 25~ns.

Reading out from Timepix3 is facilitated by an FPGA that acts as a middleman between the Timepix3
ASIC and the acquisition computer. The ``Speedy PIxel Detector Readout'' (SPIDR) readout
system~\cite{Visser:JINST10:C12028, Heijden:JINST12:C02040} provides both a 10~Gbps optical and
1~Gbps ethernet interface with the former allowing for the full 40~Mhits/cm$^{2}$/s hit rate. It
also extends the ToA timestamp range to 26~s by including an additional 16-bit timestamp at the end
of a pixel data packet. Both the ToA and SPIDR are synchronized with a global 48-bit clock. This
clock can be used to further extend the timestamp range to 81~days. SPIDR also includes an external
trigger input that introduces an additional trigger timestamp with a resolution of 260~ps allowing
for, \eg, event selection or time-of-flight (ToF) mass spectroscopy (MS) with a precise time
reference from, for instance, a pulsed laser system. With SPIDR, `slow' communication such as
configuring Timepix and uploading pixel parameters are done through the TCP whilst `fast'
communication, such as pixel and time data packets during acquisition, is handled using
UDP~\cite{Visser:JINST10:C12028}.

Python is rapidly growing to become the \emph{defacto} language in scientific applications with its
rich ecosystem of stable open-source libraries and tools that provide fast and easy data processing.
Image analysis is supported by libraries such as OpenCV~\cite{Bradski:OpenCV} and
scipy~\cite{Jones:scipy:2001:fudged} and plotting through matplotlib~\cite{Hunter:CompSciEng9:90}
either statically or interactively, \eg, using IPython~\cite{Perez:CompSciEng9:21}, means that
traditional cameras can go from hookup to acquisition to visualization to analysis with little
effort and code.

For Timepix3, the data pipeline is far more complex. Traditionally, Timepix3 must be configured for
acquisition using a separate program such as SoPhy~\cite{ASI:website} or using the original SPIDR
C++ API from NIKHEF~\cite{Heijden:JINST12:C02040} and the UDP packet-stream must then be captured to
a file. Then, a decoding script must be written -- or used from someone else –– to correctly decode
the packets. These decoded packets then require additional processing to improve time resolution. If
event selection or time referencing is required triggers must be correlated to the correct pixel
packets. Essentially a great deal of effort is required by the user to even begin visually
inspecting and analysing the data-stream. Here, we provide the \pymepix framework to bring the same
ease and usability as for traditional cameras to the use of Timepix3 with SPIDR to the scientific
community.

\section{\pymepix library}
\label{sec:pymepix-operation}
\pymepix\ is intended to bridge the gap between Timepix3 and Python. The goal of the library is to
allow a user without deep technical knowledge of Timepix3 to establish a connection, start
acquisition, and retrieve and plot pixel and timing information in as few lines of code as possible;
at the same time it provides all details to unleash the full power of Timepix3-SPIDR hardware. This
is achieved by classes that act as a black-box, handling all of the low level TCP communication and
decoding of the UDP data-stream, presenting them in a \emph{pythonic} fashion. More advanced and
lower-level control of SPIDR and Timepix3 is still available from these black-box classes or can be
established directly by the user. For easy installation, it only depends on the standard python
library, \textit{numpy}, and \textit{scikit-learn}.

One of the key features in \pymepix is a high-performance data-pipeline. The currently available
pipeline is capable of decoding all UDP packets, extending the timestamp and, if needed, correlating
triggers to pixels at a data-rate of 1 Gbps. A user can deal with
the datastream immediately, but \pymepix also offers a more convenient method that can be leveraged
with more interactive flavors of python such as \texttt{IPython} or \texttt{Jupyter Notebook}.

In many applications, \eg, the electron and ion imaging performed in our experiments, single
``physics events'' are detected by many detector pixels simultaneously. Thus, the individual-pixel
information should be merged to provide the most precise information on the actual physical event.
Rather than relegating this step to post-data acquisition, it instead motivated our need and
inclusion of an easily customizable pipeline that any user can leverage to include their own
post-processing routine, such as, \eg, centroiding or time-to-mass conversion in ion imaging,
providing a more meaningful datastream directly from 'Timepix3' during data acqusition.

Whilst originally developed to simplify data-retrieval in ion and electron imaging experiments
within the Controlled Molecule Imaging group~\cite{Chang:IRPC34:557, Kierspel:PCCP20:20205}, we
envision that the library will help scientific and industrial use of Timepix3 and allow for more
novel applications of the hardware. While we describe a slice of the capabilities provided by the
library in this introductory article, this is not comprehensive. Full documentation is available
through sphinx documentation built-in to the library code itself.

\subsection{Data model}
\label{sec:data-model}
Before looking at how the library is used, it is useful to describe how data from Timepix3 is
represented. In \pymepix, all data that is retrieved from Timepix3 is packaged in the form of a
python tuple called a \textit{message} that includes a \textit{header} and the actual data. The
header is essentially an integer that identifies what kind of data the message contains. \pymepix
aliases these headers using the \texttt{IntEnum} \texttt{MessageType}. Here's an example message
with the header \texttt{MessageType.PixelData}:
\begin{lstlisting}[frame=leftline]
(<MessageType.PixelData: 1>,
 (array([4, 4, 4, 4, 4, 4, 4, 4, 4, 4, 4, 4, 4, 4, 4, 4, 4, 4, 4],
        dtype=uint64),
  array([128, 128, 128, 128, 128, 128, 128, 128, 128, 128, 128, 128, 128,
         128, 128, 128, 128, 128, 128], dtype=uint64),
  array([0.02564228, 0.05975617, 0.06174524, 0.09620246, 0.1023057 ,
         0.19506415, 0.20917037, 0.21590329, 0.21627018, 0.22074089,
         0.2308115 , 0.23354445, 0.2339248 , 0.24488667, 0.28150781,
         0.28533644, 0.29780308, 0.33211206, 0.33664652]),
  array([   50, 25550, 25550, 25550,    25, 25550, 25550, 25550, 25550,
         25550, 25550, 25550,    50,    75, 25550,   200,    25, 25550,
         25550], dtype=uint64)))
\end{lstlisting}
Unpacking this gives us the header and the data. The data part can be further unpacked into four
separate arrays: pixel $x$ coordinate, pixel $y$ coordinate, extended pixel ToA in seconds, and ToT
in seconds. For messages with arrays, the elements of each array represent the
indices of a signal pixel. In our example, the first pixel was located at $(4,128)$ and arrived at
$\ordsim0.0256$~s. \autoref{tab:data-formats} gives a summary of the available messages, their
identifiers, and data format.

\begin{table}
   \centering%
   \begin{tabular}{lll}
     \hline\hline
     \texttt{MessageType} Identifier   & Description         & Data                            \\
     \hline
     \texttt{RawData}      & Raw UDP packets     & packets, timestamp              \\
     \texttt{PixelData}    & Decoded pixels      & $x$, $y$, global ToA, ToT           \\
     \texttt{TriggerData}  & Decoded triggers    & trigger~\#, global trigger Time \\
     \texttt{EventData}    & Correlated triggers & trigger~\#, $x$, $y$, rToA, ToT     \\
     \texttt{CentroidData} & Centroided events   & trigger~\#, $x$, $y$, rToA, ToT    \\
     \hline\hline \\
   \end{tabular}
   \caption{A summary of messages provided by \pymepix. All items under Data are in the form of
      arrays except for \texttt{timestamp}, which is an integer. The \emph{global} prefix refers to
      timestamps that have been extended past their original range. \emph{rToA} refers to the time
      of arrival computed relative to the associated trigger time. All time of arrival and trigger
      times are in seconds. See text for further details.}
   \label{tab:data-formats}
\end{table}

\subsection{Connection and Acquisition}
\pymepix\ provides the high-level class \texttt{Pymepix} that handles the connection to SPIDR,
enumerates all available Timepix devices and manages acquisition. It simply requires the TCP IP
address and port to SPIDR for instantiation:
\begin{lstlisting}[frame=leftline]
>>> from pymepix import Pymepix
>>> ppx = Pymepix(('192.168.1.10',50000))
\end{lstlisting}
\texttt{Pymepix} overwrites the \texttt{\_\_len\_\_} operator so calling \texttt{len} gives the
total number of Timepix devices connected:
\begin{lstlisting}[frame=leftline]
>>> len(ppx)
1
\end{lstlisting}
Acquisition can simply be started using the \texttt{start} method:
\begin{lstlisting}[frame=leftline]
>>> ppx.start()
\end{lstlisting}

Acquiring data comes in two flavors: polling and callbacks. Polling is a more relaxed way of
retrieving data and is the default method of acquisition. When activated, all data is placed into an
internal fixed length ring buffer array. Getting data is simply a case of calling the \texttt{poll}
method:
\begin{lstlisting}[frame=leftline]
>>> ppx.poll()
(<MessageType.PixelData: 1>,
 (array([72, 4, 120, 4], dtype=uint64),
  array([119, 20, 124, 100], dtype=uint64),
  array([0.37300809, 0.37479955, 0.38450748, 0.38831139]),
  array([25550, 25550, 25550, 25550],
        dtype=uint64)))
\end{lstlisting}
This pops the oldest value in the buffer. When new data comes in and the buffer is full, the oldest
value is overwritten. An empty buffer raises a \texttt{PollBufferEmpty} exception.

For a more immediate handling of data, callbacks can be used. In this mode as soon as a message
arrives it is split into its \textit{header} and \textit{data} parts and passed into the function
provided by the user. This mode is immediately enabled when a function is assigned to the
\texttt{dataCallback} property:
\begin{lstlisting}[label={lst:callback},frame=leftline]
>>> def my_callback(header,data):
        if header is MessageType.PixelData:
            print('Pixels')
        else:
            print('Not pixels')
>>> ppx.dataCallback = my_callback
Pixels
Not Pixels
Not Pixels
Pixels
\end{lstlisting}
Polling is disabled at this point but can be enabled by calling the \texttt{enablePolling} method.
Either methods can be switched between during acquisition. Of course, acquisition can be stopped
through the \texttt{stop} method:
\begin{lstlisting}[frame=leftline]
>>> ppx.stop()
\end{lstlisting}

\subsection{Configuring Timepix}
During initialization, \texttt{Pymepix} enumerates connected Timepix3 devices and creates a
\texttt{TimepixDevice} object for each one. These can be used to configure each individual Timepix3
device and can be accessed by the square-brackets operator:
\begin{lstlisting}[frame=leftline]
>>>tpx = ppx[0]
>>>tpx.deviceName
W0028_G05
\end{lstlisting}
With the \texttt{deviceName} property returning the assigned name of the Timepix3 device.
\texttt{TimepixDevice} gives full access to all Timepix registers and digital-to-analog-converter
(DAC) parameters, such as current preamplifier settings and voltage thresholds, in the form of class
properties. When dealing with registers, \pymepix provides a set of \texttt{enums} that can be used
to easily understand and set them. For instance, reading the operation mode register and switching
from ToA only mode to ToA-and-ToT mode is easily and clearly achieved:
\begin{lstlisting}[frame=leftline]
>>> tpx.operationMode
OperationMode.ToA
>>> tpx.operationMode = OperationMode.ToAandToT
>>> tpx.operationMode
OperationMode.ToAandToT
\end{lstlisting}
With regards to DAC parameters, it is important for previous users of Timepix3 to note that in
\pymepix\ they are now interpreted in SI units, \eg, Volt (V) and Ampere (A), rather than integer
codes. Commonly tuned DAC parameters are the coarse and fine voltage thresholds. Reading and then
setting these to 580~mV and 100~mV, respectively, is achieved through
\begin{lstlisting}[frame=leftline]
>>> tpx.Vthreshold_coarse
0.480
>>> tpx.Vthreshold_fine
75
>>> tpx.Vthreshold_fine = 100.0
>>> tpx.Vthreshold_coarse = 580.0
\end{lstlisting}
Of course if you're unsure of the units used you can always check the docstring:
\begin{lstlisting}[frame=leftline]
>>> tpx.Ibias_Ikrum.__doc__
nA
\end{lstlisting}
For backward compatibility, the option to set DAC parameters using a supplied integer code and value
is available through the \texttt{setDac} method:
\begin{lstlisting}[frame=leftline]
>>> tpx.VThreshold_coarse
0.480
>>> thresh_code,thresh_value = 7,8
>>> tpx.setDac(thresh_code,thresh_value)
>>> tpx.VThreshold_coarse
0.560
\end{lstlisting}

Pixel configuration is handled with the properties \texttt{pixelMask} for a pixel mask and
\texttt{pixelThreshold} for the pixel DAC thresholds. They are local configurations that can be
passed to and from Timepix3 using the \texttt{uploadPixels} and \texttt{refreshPixels} methods,
respectively. These are exposed as Numpy \texttt{ndarrays} of type \texttt{numpy.uint8} and can thus
be used like any other Numpy array:
\begin{lstlisting}[frame=leftline]
>>> tpx.pixelMask[...] = 1
>>> tpx.pixelThreshold[::2,1::2] = 8
\end{lstlisting}
This includes being used as arguments whereever Numpy arrays are accepted
\begin{lstlisting}[frame=leftline]
>>> numpy.fill_diagonal(tpx.pixelMask,0)
>>> import matplotlib.pyplot as plt
>>> plt.matshow(tpx.pixelMask)
\end{lstlisting}
and new arrays can be assigned to them as long as they are of shape $256\times256$ with type
\texttt{numpy.uint8}:
\begin{lstlisting}[frame=leftline]
>>> tpx.pixelThreshold = np.ones(shape=(256,256),dtype=np.uint8)
\end{lstlisting}
Assigning an array with an incorrect shape and type results in a \texttt{BadPixelFormat} exception.
\texttt{TimepixDevice} also handles loading settings and the pixel matrix from configuration files
using the \texttt{loadConfig} method. The behaviour of \texttt{loadConfig} is determined by what
\texttt{TimepixConfig} class is assigned. This class simply provides methods \texttt{dacCodes},
\texttt{maskConfig}, and \texttt{thresholdConfig}, which provide a list of DAC code-value pairs, the
pixel mask array, and the pixel threshold array, respectively. Therefore, custom file formats can be
supported by overriding these methods in your own class and passing it into the
\texttt{setConfigClass} method. The arguments of \texttt{loadConfig} essentially become the
initialization arguments of the class. One of the simplest implemented is the \texttt{DefaultConfig}
class which simply sets all DAC parameters to their default values and clears all pixel
configuration matrices:
\begin{lstlisting}[frame=leftline]
>>> tpx.setConfigClass(DefaultConfig)
>>> tpx.loadConfig()
\end{lstlisting}
Another is \texttt{SophyConfig} which loads DAC parameters and pixel matrices from Sophy \mbox{.spx}
files
\begin{lstlisting}[frame=leftline]
>>> tpx.setConfigClass(SophyConfig)
>>> tpx.loadConfig('W0028_G05_50V.spx')
>>> tpx.loadConfig('W0028_G05_30V.spx')
\end{lstlisting}
This is also the default configuration class set on \texttt{TimepixDevice} initialization.

\subsection{Data processing pipeline}
\label{sec:dataprocessing}
Embedded within each \tpxdevice object is a parallel data-pipeline that performs the data-processing
for the respective Timepix device. The output of the pipeline, and by extension the callbacks and
\texttt{poll} method of \texttt{Pymepix}, is determined by assigning an appropriate \acqpipeline
class through the \texttt{setupAcquisition} method.
\begin{lstlisting}[frame=leftline]
>>> tpx.setupAcquisition(PixelPipeline)
\end{lstlisting}
\sloppy%
The \acqpipeline can be thought of as a recipe on how to build the data-pipeline using \pipeobj{s},
which handle the actual processing. Each \pipeobj is in-fact a Python
\texttt{multiprocessing.Process} with a \texttt{process} method that can generate or perform work on
any piece of data in the form of messages. Therefore, the \acqpipeline builds and connects up all
the necessary \pipeobj in the correct order, spawns their processes and manages the flow of data
through the pipeline. The pipeline is built each time the \texttt{start} method is called in
\texttt{Pymepix} and is destroyed by \texttt{stop}. Some predefined \acqpipeline{s} include the
\texttt{PixelPipeline} that simply reads from UDP and decodes pixels and triggers or the
\texttt{CentroidPipeline} that adds a centroiding stage. The currently set \acqpipeline is accessed
through the \texttt{acquisition} property in \tpxdevice. For example, in event mode,
\texttt{PixelPipeline} can enable pixel and trigger correlation through its \texttt{enableEvents}
property:
\begin{lstlisting}[frame=leftline]
>>> tpx.acquisition.enableEvents
False
>>> tpx.acquisition.enableEvents = True
\end{lstlisting}
These pipelines can be augmented further by attaching your own \pipeobj. The following code snippet
demonstrates a simple custom concrete \pipeobj :
\begin{lstlisting}[frame=leftline]
class FooBar(BasePipelineObject):
    FooData = 25
    def __init__(self,input_queue=None,output_queue=None):
        super().__init__(input_queue,output_queue)

    def process(self,header,data):
        if header is MessageType.PixelData:
            self.pushOutput(self.FooData,'Bar')
\end{lstlisting}
The \texttt{input\_queue} and \texttt{output\_queue} are the input and output
\texttt{multiprocessing.Queue}{s}, respectively, that handle movement of data. A user only needs to
know that they are required for the class to function, but do not have to interact with it. In the
code above, we see that the \texttt{process} method accepts a message header and message data as
parameters, similar to callbacks in \autoref{lst:callback}, with the movement of the result to the
next stage carried out by \texttt{pushOutput} using a unique message type integer identifier
\texttt{FooData}. Attaching it to the pipeline can be done using the \texttt{addStage} method:
\begin{lstlisting}[frame=leftline]
>>> tpx.acquisition.addStage(10,FooBar)
\end{lstlisting}
The first argument is the stage number and define what order the messages flow through each pipeline
object in the pipeline. Lower numbers are earlier in the pipeline. Starting an acquisition now shows
the new message from our class:
\begin{lstlisting}[frame=leftline]
>>> ppx.start()
>>> ppx.poll()
(25, 'Bar')
\end{lstlisting}
Some of the main specific \pipeobj classes defined in \pymepix are briefly described in the
following:

\paragraph{UdpSampler}
This pipeline object creates a UDP socket connection to Timepix3 and immediately begins collecting
UDP packets. UDP packets are converted into arrays and timestamped before being sent off to the next
stage with the \texttt{MessageType.RawData} identifier.

\paragraph{PacketProcessor}
\texttt{PacketProcessor} processes messages with the header \texttt{MessageType.RawData}. It has two
modes of operation, pixel mode and event mode. In pixel mode, each 64-bit pixel packet is decoded
into their representative column, row, time of arrival (ToA), time over threshold (ToT) and SPIDR
timestamp. The time of arrival timestamp range is first extended from 407 \us\ to 26.8 s using the
SPIDR timestamp. Using the attached 48-bit global timestamp from the \texttt{UdpSampler} message,
the range of unambiguity is further extended to 81.4 days. If fast time of arrival is enabled, then
this is applied to the timestamp improving its resolution from 25~ns to 1.56~ns. Finally, the
timestamp is converted from integer nanoseconds into double precision seconds. The results are
pushed into the output queue with the \texttt{MessageType.PixelData} Any triggers encountered are
also decoded and their timestamp extended in the same fashion as the pixel. The output has the
\texttt{MessageType.TriggerData} identifier.

Event mode follows the same procedure as pixel mode but here the trigger and pixel data are cached.
When a certain number of triggers have been collected, the cached triggers are then used as bins to
assign the pixel ToA to their appropriate triggers using the \texttt{numpy.digitize} method. Each
pixel has an assigned trigger number and their ToA are computed as relative to their assigned
triggers time. The output has the identifier \texttt{MessageType.EventData}.

\paragraph{Centroiding}
\texttt{Centroiding} works on \texttt{MessageType.EventData} messages to improve spatial and
temporal resolution by centroiding. Since we are working with three-dimensional data
$(x,y,\text{ToF})$, flood fill algorithms are unsuitable as they risk destroying events that occur
later within the same pixel. Instead the DBSCAN~\cite{Ester:PSICKDDM:226} algorithm within the
\texttt{scikit-learn}~\cite{Pedregosa:JMLR12:2825} package was chosen as most other clustering
algorithms either require \emph{a~priori} knowledge of the number of clusters or were not fast
enough. For our dataset, the euclidean metric is used, \ie, the $\epsilon$ term represents the
maximum distance radius that points should be considered as part of a cluster. Since the event ToA
operates on a different scale, it is necessary to introduce an extra $\epsilon_f$ term that is the
maximum time distance between pixels. The event ToA are scaled as
$t_f'=\epsilon/\epsilon_f\cdot{}t_f$, where $t_f$ is the original event time and $t_{f}'$ is the
scaled event ToA. The default values used are empirically chosen as $\epsilon=3$~ns and
$\epsilon_f=500$~ns as these values give good results in our analysis. Once the clusters are
established, the centroided $x$, $y$, and event time are computed as a weighted mean of all points
with respect to their ToT. The centroids final ToT is the maximum ToT in the cluster. The output is
packaged under the \texttt{MessageType.CentroidData}.

\subsection{Timewalk Correction}
Timewalk is an often discussed effect in the literature involving Timepix3 and improves the temporal
resolution by calibrating out a dependency of the ToA on the ToT. The standard operating
procedure~\cite{Turecek:JINST12:C12065} is to first calibrate the pixels. This calibration is well
known, \eg, already from Timepix detectors~\cite{Jakubek:NIMA633:S262}. To assist with this the
\texttt{pymepix.util} module provides the helper function \texttt{generate\char`_ timewalk\char`_
   lookup} that generates a timewalk lookup array for a given ToA/ToF region and ToA/ToF and ToT
array. Time arrays from the \texttt{MessageType.PixelData}, \texttt{MessageType.EventData}, and
\texttt{MessageType.CentroidData} can be used. The function works similarly to the methods in
literature~\cite{Turecek:JINST12:C12065, Zhao:RSI88:113104} by first assuming a ``true'' arrival
time computed from averaging the ToA within a narrow slice of the highest ToTs within the region.
All ToAs are converted into time difference from the ``true'' ToA, are sampled for each ToT, and fit
to a Gaussian whose expectation value $\mu$ is the timewalk value. Since the ToT is a 10~bit value,
there are only 1024 possible values it can take. The lookup table is generated for all possible ToT
values. The index $i$ for a given ToT $t$ in nanosecond can be easily computed as
\begin{equation}
   i = \frac{t}{25} - 1
\end{equation}
The resulting array can be stored or used by the user to correct pixel or event data after
acquisition.

\section{Enduser utility programs}
Included with pymepix are two programs, \pymepixacq and \pymepixview, that allow for the immediate
connection and acquisition of Timepix3 ``out of the box''.

\subsection{Command-line data acquisition with \pymepixacq}
\label{sec:pymepixacq}
\pymepixacq is a command line code using the \pymepix library to acquire from a single Timepix
device. Its help output is:
\begin{lstlisting}[label={lst:pymepixacq}, frame=leftline, language=bash, keywords={}]
> pymepix-acq -h
usage: pymepix-acq [-h] [-i IP] [-p PORT] [-s SPX] [-v BIAS] -t TIME -o OUTPUT
                   [-d DECODE] [-T TOF]

Timepix acquisition script

optional arguments:
-h, --help                 show this help message and exit
-i IP, --ip IP             IP address of Timepix
-p PORT, --port PORT       TCP port to use for the connection
-s SPX, --spx SPX          Sophy config file to load
-v BIAS, --bias BIAS       Bias voltage in Volts
-t TIME, --time TIME       Acquisition time in seconds
-o OUTPUT, --output OUTPUT output filename prefix
-d DECODE, --decode DECODE Store decoded values instead
-T TOF, --tof TOF          Compute TOF if decode is enabled
\end{lstlisting}
The application can be used from the command-line by specifying an output filename and time in
seconds:
\begin{lstlisting}[frame=leftline]
pymepix-acq --time 10 --output my_file
\end{lstlisting}
by default only UDP packets are stored with the \texttt{.raw} file extension in raw byte format
which includes the 48-bit Timepix3 clock packets. To store decoded pixel values the \texttt{-d}
switch can be used:
\begin{lstlisting}[frame=leftline]
pymepix-acq -d --time 10 --output my_file
\end{lstlisting}
this stores the pixel data format specified in \autoref{tab:data-formats} through successive
\texttt{numpy.save} statements into a single file with the \texttt{.toa} file extension. Loading the
data in an user program is achieved by repeatedly executing four \texttt{numpy.load}, for
$x$, $y$, global ToA, and ToT, respectively, until an end of file exception occurs.
Adding the additional \texttt{-T} switch will activate the event selection mode and will store ToF
data in a file with the \texttt{.tof} file extension. This is similar to the previous format, but an
additional numpy load is required for the trigger number array.

\subsection{Graphical user interface \pymepixview}
\begin{figure}[h]
   \includegraphics[width=\linewidth]{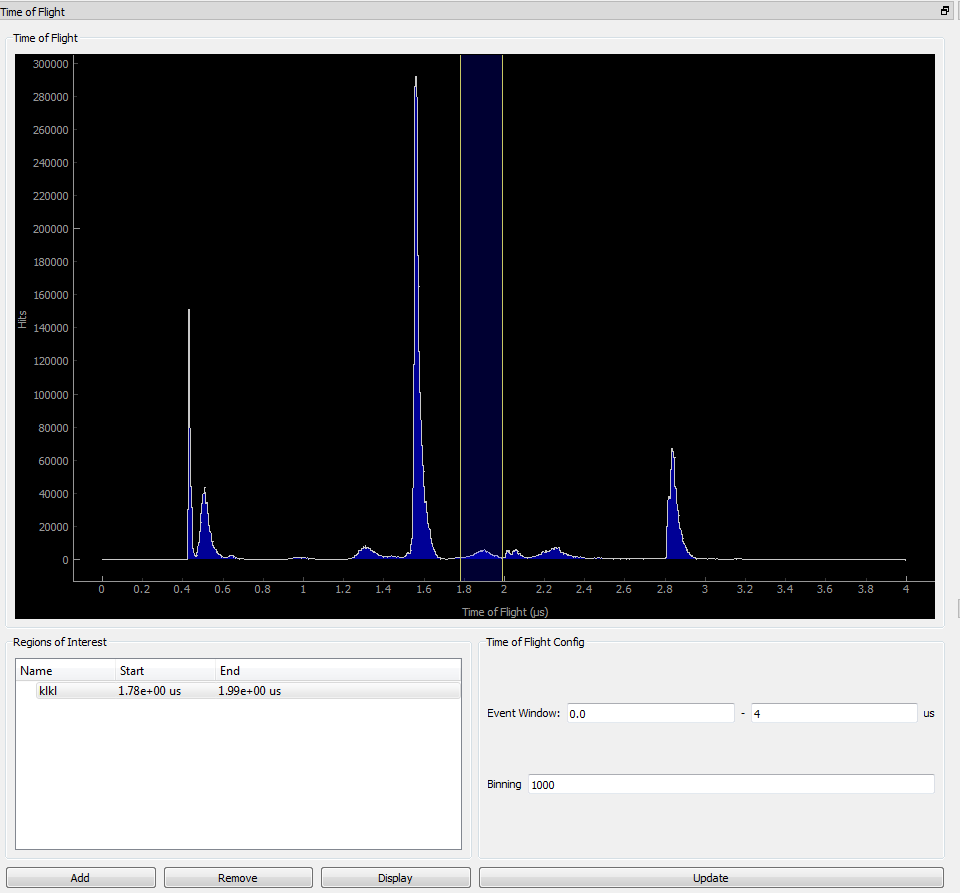}
   \caption{A section of the \pymepixview GUI displaying a time of flight spectrum
     and gate creation and selection. The shaded region corresponds the to gate defined
     in the table}
   \label{fig:daqtof}
\end{figure}

\begin{figure}[h]
   \includegraphics[width=\linewidth]{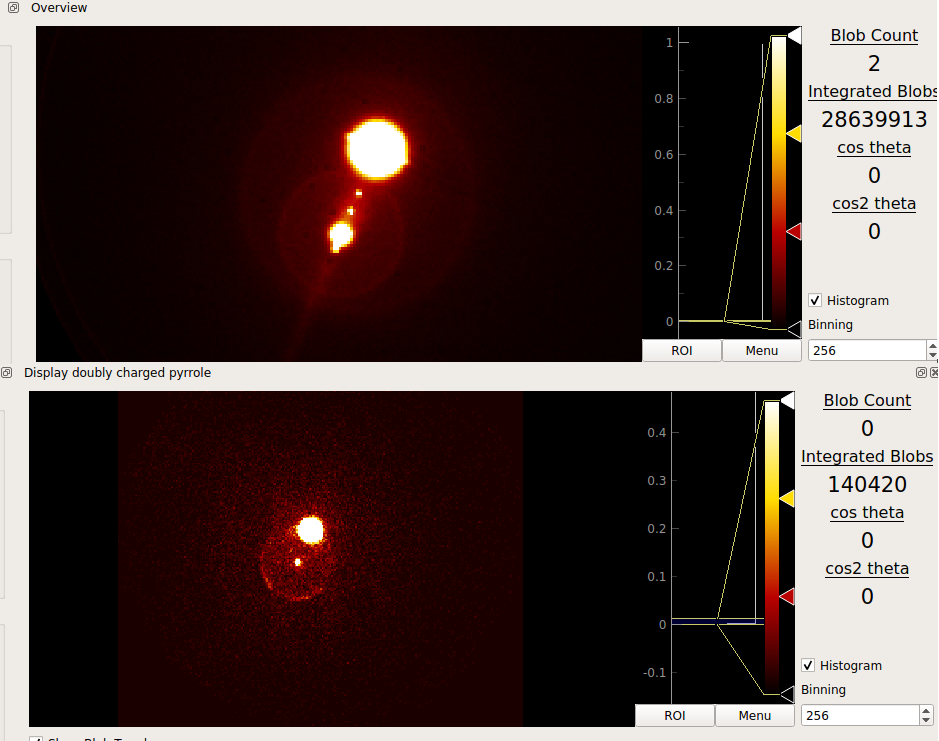}
   \caption{This figure highlights the gating feature of \pymepixview where the top plot
     is the full image seen from timepix and the bottom is a ``gated'' image created by the user.}
   \label{fig:pyrrole}
\end{figure}

A separate online viewer is included, built using the \pymepix, pyqtgraph and PyQt5 libraries.
\pymepixview acts as both a demonstration of the \pymepix library and as a simple yet
fully-fledged acquisition program. \pymepixview is geared towards ion imaging, in particularVMI
experiments~\cite{Eppink:RSI68:3477, Trippel:MP111:1738}, but it is still capable of general
use.
\begin{figure}[b]
   \includegraphics[width=\linewidth]{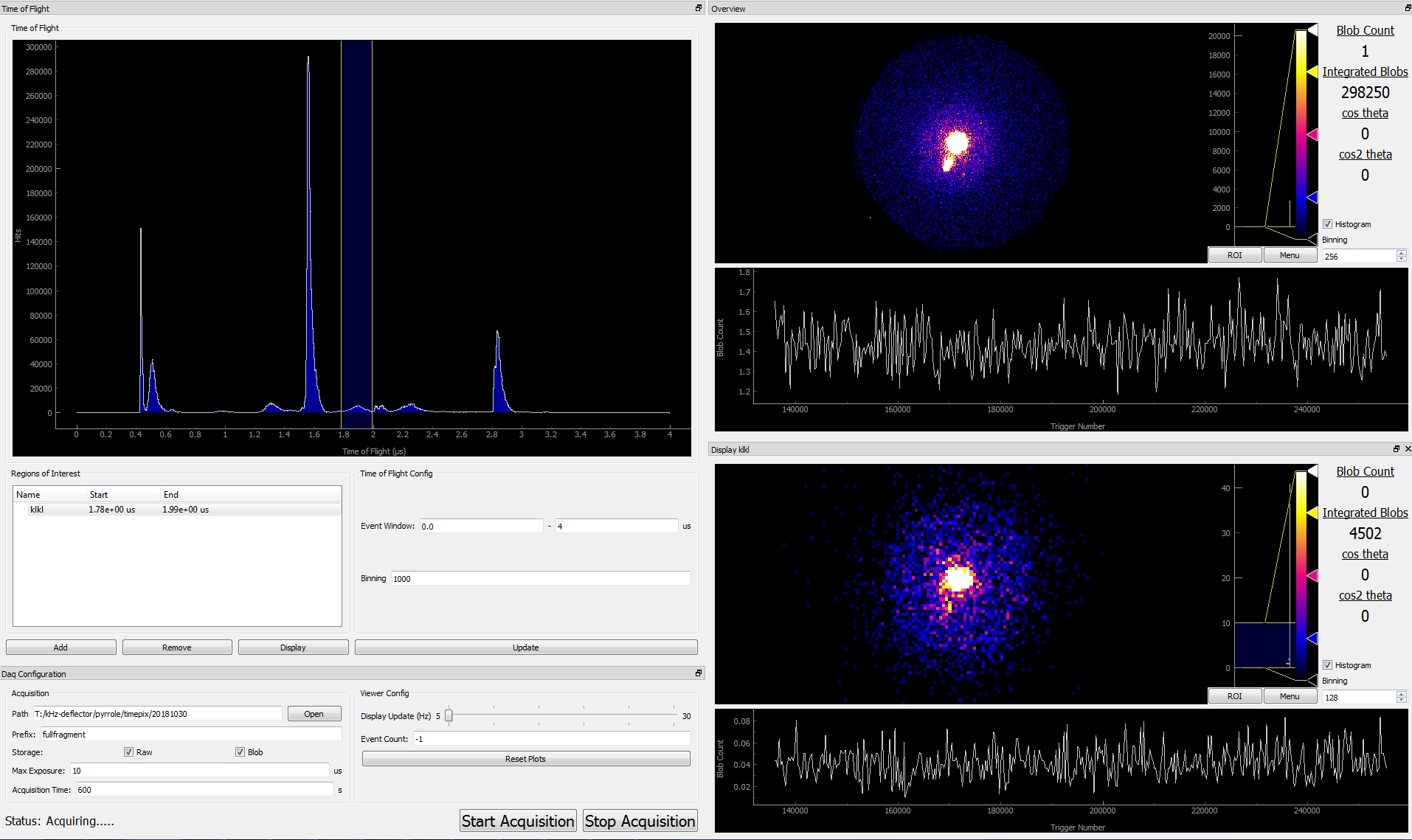}
   \caption{A screenshot of the \pymepixview GUI in operation. Shown are a live ToF spectrum in the
      upper left including a ToF gate as blue-shaded area, centroiding, ToF gating, and histograms
      over the full data in the upper right and the data from the gate in the ToF spectrum.}
   \label{fig:pymepix_daq}
\end{figure}
The viewer can acquire and store all formats described in \autoref{sec:pymepixacq}, but with the
addition of the centroided data given in \autoref{tab:data-formats} as \texttt{.blob} files. There
are three modes available in the viewer: 'ToA mode', 'ToF mode', and 'centroiding mode'. In ToA
mode, the viewer can display an integrated image over a specified time range. In ToF mode, a time of
flight spectrum is also displayed in \autoref{fig:daqtof}. 
The ToF spectrum can be used to define a number of time-ranges, so-called ``gates'', and labelled 
by the user to generate new plots in the
viewer that display and update data from the selected gates as pixel images. This can be seen in
\autoref{fig:pyrrole}where the top plot is the original ``image'' from the Timepix3 detector and 
the bottom is a gated plot created by the user. 
Centroiding mode provides the same features as ToF mode, but with the addition of
using the higher spatial and temporal resolution of the reduced centroided data in both the ToF
spectrum and the pixel plots. An additional plot can be enabled underneath each pixel plot that
shows the number of clusters, so-called ``blobs'', recorded at each event/trigger number, which can
be used for experimental parameter optimization. All of these modes can be switched between
on-the-fly.
\begin{figure}[h]
   \includegraphics[width=\linewidth]{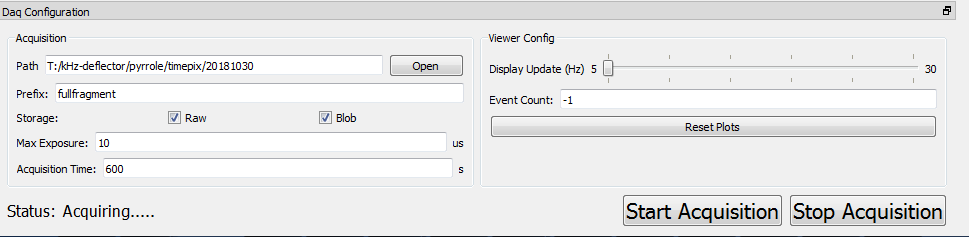}
   \caption{A section of the \pymepixview GUI highlighting acquisition control
         and storage settings}
   \label{fig:daqcontrol}
\end{figure}
\autoref{fig:daqcontrol} displays the acquisiton control widget which allows the user
to set storage location, what is stored, acquisition time and to begin the acquisition itself.

\section{Conclusion}
Timepix3 is a high-performance high-throughput time-resolving pixelated detector. Python is growing
to become a standard language in scientific applications. Connecting these worlds, \pymepix is an
open-source python library for communicating with and acquiring data from Timepix3, using SPIDR, and
simplifies the hardware's control, acquisition, and the analysis of its data. It uses a
human-understandable abstraction of the hardware whilst also allowing for fine-grained control. We
described an extendible data pipeline to provide a simple way of retrieving and processing the
Timepix3 datastream whilst maintaining performance. Included are programs that serve to not only
demonstrate the ease of use of \pymepix, but to also allow the community to rapidly understand and
make use of the powerful Timepix3 detector.

We point out that in applications such as, for instance, ion or electron imaging a detector system
built from Timepix3, SPIDR, and \pymepix, introduced here, can easily enable single-shot acquisition
of images at very high repetition rate, such as at current or existing free-electron laser
facilities~\cite{Feldhaus:JPB43:194002, Abela:DESY:2006, SLAC:New-Science-LCLS-II:2016} or
high-repetition rate table-top laser systems~\cite{Rothhardt:OptExp24:18133, Furch:OL42:2495}.

\section*{Program Availablity}
\label{sec:program-availablity}
The library and programs are available under the GPLv3 license from a git repository at
\url{https://stash.desy.de/scm/cmipublic/timepix.git} and the Python pip repository.

\section*{Acknowledgments}
We gratefully acknowledge helpful discussions with Heinz Graafsma regarding Timepix3 capabilities
and use. Igor Rubinskiy had originally started work with Timepix3 in our group and Ruth Livingstone
and Milija Sarajlic had contributed to early experiments. Some low-level codes for the programming
of SPIDR were obtained from the C++ code developed by Henk Boterenbrood and colleagues at Nikhef and
Amsterdam Scientific Instruments and we gratefully appreciate the sharing of these with us.

This work has been supported by the Clusters of Excellence ``Center for Ultrafast Imaging'' (CUI,
EXC~1074, ID~194651731) and ``Advanced Imaging of Matter'' (AIM, EXC~2056, ID~390715994) of the
Deutsche Forschungsgemeinschaft (DFG) and by the European Research Council under the European
Union's Seventh Framework Program (FP7/2007-2013) through the Consolidator Grant COMOTION
(ERC-Küpper-614507) and the European Union's Horizon 2020 research and innovation program under the
Marie Skłodowska-Curie Grant Agreement 641789 ``Molecular Electron Dynamics investigated by Intense
Fields and Attosecond Pulses'' (MEDEA).

\bibliographystyle{JHEP}

\providecommand{\href}[2]{#2}\begingroup\raggedright\endgroup
   
\end{document}